# When Does Tourism Raise Land Prices?
# Threshold Effects, Superstar Cities, and Policy Lessons from Japan

Mingzhi Xiao[1]*, Takara Sakai[1], Daisuke Murakami[2], Yuki Takayama[1]

**Abstract:** While tourism is widely regarded as a catalyst for economic and urban transformation, its effects on land prices remain contested. This study investigates the relationship between tourism and land prices using a comprehensive panel of 1,724 Japanese municipalities from 2021 to 2024, with annual tourist arrivals as a proxy for tourism activity. Employing mediation analysis and panel threshold regression, the analysis reveals that substantial land price increases are concentrated in a small group of "superstar" cities, specifically those ranking in the top 5.9% for tourist arrivals, while most municipalities experience little or no effect. The findings highlight pronounced nonlinearities and spatial heterogeneity in tourism's economic impact across Japan. The potential mechanisms linking tourism to land price growth are mixed, with possible benefits for local residents as well as risks of increased burdens. These results underscore the need for policies that promote inclusive growth and equitable distribution of tourism-related gains.

## 1. Introduction

Tourism has become one of the major engines of economic growth and urban transformation worldwide. According to the United Nations World Tourism Organization, international tourist arrivals surpassed 1.5 billion and generated over $1.4 trillion in export revenues globally in 2019, making tourism not only a vital contributor to national GDP but also a fundamental force reshaping cities, labor markets, and social structures (Pulido-Fernández & Cárdenas-García, 2021). In many countries, the expansion of tourism has driven job creation, entrepreneurial activity, infrastructure investment, and population mobility, thereby fueling both urban renewal and regional development (Dana et al., 2014; Biagi et al., 2015).

Yet, alongside these benefits, the rapid growth of tourism has intensified debate about its broader social and spatial consequences. Researchers and policymakers increasingly question the impact of tourism on local housing and land markets, as increasing tourist arrivals have been linked to affordability pressures, neighborhood transformation, and gentrification in cities around the world (Mikulić et al., 2021; Gotham, 2013; Zhang, 2024). Notably, such dynamics are especially pronounced in East Asia, where rapid economic growth, policy-driven tourism promotion, and demographic change have intersected to reshape urban and regional landscapes (Song et al., 2024; Lv, 2019).

---
[1] Institute of Science Tokyo, 2-12-1 W6-9, Ookayama, Meguro-ku, Tokyo 152-8552, Japan
[2] Institute of Statistical Mathematics, 10-3 Midori-cho, Tachikawa, Tokyo 190-8562, Japan
* Corresponding author: xiao.m.1475@m.isct.ac.jp

Within East Asia, Japan stands out as a particularly revealing case. By 2019, Japan ranked 11th globally and 3rd in Asia for international tourist arrivals, hosting nearly 32 million inbound visitors and earning over $46 billion in international tourism receipts. Domestic tourism plays an even greater role, with revenues exceeding ¥21.9 trillion (approx. $200 billion), firmly establishing tourism as a core pillar of the country's "Regional Revitalization" strategy. Policymakers have thus positioned tourism as both an economic catalyst and a tool to address urban–rural disparities, population decline, and local innovation (Mahadevan & Suardi, 2019).

However, the surge in tourism has also brought substantial challenges. In iconic cities such as Kyoto, a dramatic influx of visitors has coincided with steep increases in land and housing prices, heightened gentrification pressures, and the displacement of long-term residents, particularly younger families (Yoshida & Kato, 2024; Gotham, 2013). Similar trends are evident in other global tourist hotspots, including Barcelona, Venice, and Dubrovnik, where the proliferation of short-term rentals (e.g., Airbnb) has altered neighborhood dynamics, eroded affordability, and prompted calls for policy intervention (Mikulić et al., 2021; Vizek et al., 2024). Even in China, rapid tourism-driven investment has contributed to soaring land prices and urban restructuring in major cities and western regions (Song et al., 2024; Li et al., 2022).

Despite extensive academic and policy attention, there is still surprisingly limited understanding of when and how tourism affects land prices, especially at the national scale and across diverse urban contexts (Cró & Martins, 2024; Chiu & Yeh, 2017; Paramati & Roca, 2019). Most existing research, both in Japan and internationally, focuses on prominent destinations or individual cities, typically relying on case studies or data from limited time periods. This leaves several fundamental questions unresolved: Does tourism invariably raise land prices? Are positive effects confined to "superstar" destinations?

To address these critical gaps, this study constructs a comprehensive panel dataset encompassing all 1,724 municipalities in Japan from 2021 to 2024. Leveraging advanced econometric techniques, including panel threshold regression (Hansen, 1999) and mediation analysis, we systematically examine: (1) the conditions under which changes in tourist arrivals influence land prices; (2) the nonlinear and heterogeneous effects of tourist arrivals on land prices (Biagi et al., 2015); and (3) the main mechanisms through which tourist arrivals, particularly via the expansion of the service sector, affect land prices (Chao et al., 2009; Pulido-Fernández & Cárdenas-García, 2021). Our analysis uncovers a pronounced threshold effect: only municipalities with exceptionally high levels of tourism exhibit significant land price appreciation, while most experience little or no change. Furthermore, we show that the expansion of accommodation and food service sectors is the primary channel through which tourism raises land prices.

Importantly, these findings fundamentally challenge the prevailing assumption that tourism universally raises land prices and always threatens affordability. Instead, our evidence shows that the scale, direction, and welfare implications of price changes are highly context dependent. Rising land prices are not inherently negative. They may signal local economic vitality and improved prosperity when accompanied by broad-based income growth but can also create additional burdens if gains are narrowly distributed. Disentangling these contrasting scenarios is crucial for the development of more targeted, equitable, and sustainable urban policies.

The remainder of this paper is organized as follows. Section 2 reviews the international and Japanese literature on the impact of tourism on land prices, highlighting theoretical and empirical gaps addressed in this study. Section 3 presents the conceptual framework and describes the data sources and variables used in the analysis. Section 4 outlines the empirical methodology, including baseline regression, mediation analysis, threshold models, heterogeneity analysis, and robustness checks. Section 5 reports the main empirical results, focusing on the threshold and mediation effects, spatial heterogeneity, and the robustness of findings. Section 6 discusses the theoretical and policy implications of the results. Section 7 provides targeted policy recommendations for local and national governments. Finally, Section 8 concludes by summarizing the key contributions and identifying directions for future research.

## 2. Literature Review

The impact of tourism on land prices has become a central focus of international research in urban studies, regional science, and applied economics. Classic theoretical land use frameworks (Alonso, 1964; Fujita, 1989; Duranton & Puga, 2015) posit that increased demand for proximity to amenities, including tourist attractions, raises the willingness to pay for land, thereby driving up prices and triggering spatial restructuring. Building on these foundations, more recent empirical work extends these ideas, arguing that tourism can indirectly stimulate property markets by generating investment in infrastructure, extending industrial supply chains, and encouraging agglomeration economies (Biagi et al., 2015; Pulido-Fernández & Cárdenas-García, 2021). Furthermore, dynamic general equilibrium models, such as Chao et al. (2009), illustrate how tourism growth can simultaneously affect local employment, welfare, and land values over time, depending on broader economic linkages.

A large body of cross-national evidence confirms that tourism growth generally pushes up urban land and housing prices, though the magnitude and distribution of these effects vary widely across contexts. In Europe, for example, Mikulić et al. (2021) find that the rapid expansion of tourism in Croatia significantly worsened local housing affordability, especially during peak seasons—a phenomenon known as the "negative externalities" of overtourism. Vizek et al. (2024) further show that the impact of tourism on property prices is highly dependent on the type of accommodation: hotels and campgrounds tend to push up prices more than short-term rentals, which can sometimes reduce long-term rental demand. Similar patterns are observed in Spain and Italy, where the proliferation of platforms like Airbnb has driven up core city land prices, displaced long-term residents, and destabilized rental markets. Turning to Asia, recent research by Zhang (2024) highlights the dynamic interplay between tourism development and housing prices in Asian cities, emphasizing that these effects can be both direct and indirect, and are highly sensitive to local market structure and policy environment.

Recent scholarship has also drawn attention to the processes of tourism gentrification and resident displacement, which are increasingly central to international and Japanese debates. Gotham (2005) conceptualized tourism gentrification as the transformation of urban neighborhoods by tourism-driven capital and consumption, leading to rising land values, the repurposing of residential spaces for tourism use, and ultimately the displacement of vulnerable or long-term residents. The Kyoto case (Yoshida & Kato, 2024) provides empirical support for these concerns in Japan, showing that new hotel construction and surging tourist inflows are associated with heightened housing affordability risk and the out-migration of lower-income groups. However, most large-sample quantitative studies, both in Japan and internationally, have limited capacity to directly observe these displacement mechanisms due to the lack of detailed migration and household-level data. As a result, much of the current understanding relies on inference from price trends and sectoral changes rather than direct evidence on resident outcomes.

In East Asia, China provides further evidence of tourism's powerful influence on land markets. Song et al. (2024), using urban panel data and advanced econometric techniques, find that tourism-driven improvements in public services, infrastructure, and environmental quality significantly raise urban land prices, with the strongest effects observed in major tourist cities and designated tourism zones. This research also highlights the critical role of mediation channels, particularly the expansion of the accommodation and food service sectors, in transmitting the impact of tourism to land prices. Meanwhile, Japanese scholarship has contributed important case-based insights but remains more limited in national scope. For instance, Yoshida and Kato (2024) show that each new hotel in Kyoto increases nearby residential property prices by an average of ¥2 million, intensifying the financial burden on local families and echoing international debates about gentrification and displacement. Hiramatsu (2023) employs a computable general equilibrium (CGE) model to show that the benefits of tourism-driven service sector expansion accrue primarily to large cities, with peripheral and rural regions often left behind due to population outflows and resource constraints. However, most Japanese studies are confined to single cities or limited temporal windows, rarely encompassing all municipalities or rigorously uncovering underlying causal mechanisms.

A further critical theme emerging in the literature is the nonlinearity and threshold effects of tourism's impact on land prices. Butler (1980) first introduced the concept of a tourism area life cycle (TALC), arguing that the effects of tourism development are fundamentally stage-dependent and nonlinear. Extending this perspective, McKercher (1999) conceptualizes tourism systems as inherently nonlinear and chaotic, suggesting that substantial positive economic and land value impacts only occur once destinations surpass certain critical thresholds of tourism activity or infrastructure capacity. Moderate or early-stage development, by contrast, often produces limited or negligible impacts. Recent empirical studies employing dynamic threshold and regime-switching models reinforce this view: "superstar" destinations with exceptional levels of tourist inflows experience disproportionately large effects on land prices (Li et al., 2022; Vuković et al., 2023), while most places may see little change or even negative spillovers. These findings suggest that tourism's impact is highly contingent on reaching a scale where agglomeration economies or infrastructure bottlenecks begin to matter, a pattern that resonates with both the TALC model and broader theories of urban development.

The mechanisms through which tourism influences land markets have also received growing attention. The expansion of the accommodation and food service sectors—driven by rising tourism demand—is recognized as a major transmission channel (Biagi et al., 2015; Khanal et al., 2022). This sectoral growth not only stimulates job creation and local incomes but also increases competition for residential space, potentially accelerating housing price growth and affordability challenges (Mikulić et al., 2021; Vizek et al., 2024). In superstar cities, these processes can interact to intensify gentrification dynamics, while in less developed or peripheral municipalities, the scale of service sector expansion may be insufficient to meaningfully impact land values (Hiramatsu, 2023). Yet, the absence of migration, rental, or transaction-level data means that much of the evidence remains indirect, highlighting the importance of future micro-level research on displacement and housing market segmentation.

Spatial heterogeneity is increasingly recognized as another key feature of the tourism–land price nexus. The elasticity of land prices to tourism can differ substantially by city size, economic profile, and local industrial composition (Cró & Martins, 2024; Paramati & Roca, 2019). Notably, tourism-driven land price appreciation is often concentrated in a few major cities or well-known tourist centers, while most small or peripheral municipalities remain largely unaffected. This uneven distribution not only underlines the "superstar" effect but also raises critical policy challenges for managing regional inequality and urban sustainability.

On the methodological front, the field has progressed from descriptive case studies to more rigorous causal identification, employing panel data models with fixed effects, instrumental variables, mediation analysis, and threshold models to disentangle nonlinear relationships. However, endogeneity remains a persistent concern—rising land prices may themselves attract tourism investment or policy support, confounding the direction of causality (Pulido-Fernández & Cárdenas-García, 2021). While recent studies have attempted to address these issues using historical or geographic instruments (such as the density of heritage sites or hot spring resources), or by including lagged explanatory variables and placebo tests (Cong et al., 2025; Song et al., 2024), fully exogenous identification is often constrained by data limitations. Furthermore, most research lacks detailed information on intra-city migration, housing tenure, or transaction-level attributes, making it difficult to empirically disentangle displacement and gentrification processes.

In summary, the international and Japanese literatures highlight three main gaps: (1) the need for systematic, national-scale evidence on nonlinear, threshold, and heterogeneous effects of tourism on land prices; (2) the importance of unpacking the mediating role of the service sector and the mechanisms of tourism gentrification and displacement; and (3) persistent challenges in causal identification and micro-level measurement. This study directly addresses these gaps by integrating a comprehensive municipal panel, panel threshold modeling, and mediation analysis to clarify when and how tourism influences land markets in Japan. Nevertheless, due to current data constraints, we are unable to directly track resident displacement or intra-city mobility and instead focus on macro-level price dynamics and their policy implications.

# 3. Theoretical Framework and Data

## 3.1 Theoretical Framework and Hypotheses

Based on the literature reviewed above, the conceptual framework guiding this study (Figure 1) integrates classic theories from urban economics with recent empirical evidence on the impact of tourism to land markets in Japan. As a theoretical foundation, the framework draws on the bid-rent theory (Alonso, 1964; Fujita, 1989, Duranton & Puga, 2015), which argue that increased demand for proximity to amenities, including tourist attractions, raises willingness to pay for land, thus driving up prices and triggering spatial restructuring. Moreover, recent studies support the view that tourism can indirectly stimulate local property markets by encouraging investment in infrastructure, extending industrial supply chains, and promoting agglomeration economies (Biagi et al., 2015; Pulido-Fernández & Cárdenas-García, 2021; Chao et al., 2009).

In the context of Japanese municipalities, we posit that tourist arrivals affect land prices via multiple interconnected pathways. First, increased tourism directly raises demand for commercial and residential land near destinations, consistent with the bid-rent framework (Alonso, 1964). Second, tourism expansion fuels growth in the accommodation and food service sectors, which generate jobs and positive local income spillovers, further increasing demand for property (Khanal et al., 2022; Južnik Rotar et al., 2023). Third, higher tourist inflows prompt investments in public goods, amenities, and urban infrastructure. These actions not only benefit visitors but also make municipalities more attractive to residents and investors (Song et al., 2024).

Accumulating evidence indicates that these impacts are neither uniform nor linear. Both theoretical arguments and empirical studies highlight strong nonlinearities and threshold effects: significant positive impacts on land prices emerge only after tourist activity surpasses a critical scale (Li et al., 2022; Mirilovi et al., 2023). Below this threshold, marginal effects may be weak or insignificant, and the elasticity of land prices to tourism can differ widely depending on city size, economic structure, and regional context (Hiramatsu, 2023; Cró & Martins, 2024; Paramati & Roca, 2019).

Building on these insights, we articulate the following hypotheses for empirical testing:

**H1:** Tourist arrivals affect municipal land prices.

**H2:** The effect of tourist arrivals on land prices is mediated by the expansion of the accommodation and food service sectors.

**H3:** The tourism–land price relationship is nonlinear, with effects concentrated above an empirically estimated tourism threshold.

**H4:** The elasticity of land prices with respect to tourist arrivals varies by municipal population size (and, where applicable, by region).

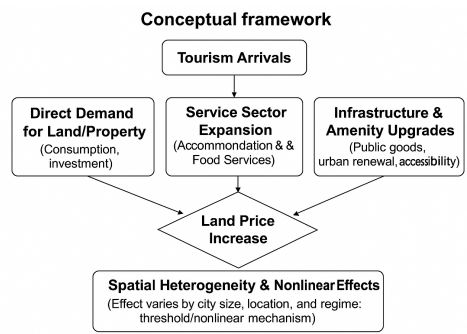

**Figure 1. Theoretical Framework**

## 3.2 Data and Variables

This study uses an annual panel dataset covering all Japanese municipalities (Si, Ku, Cho, Son) from 2021 to 2024. The key variables and data sources are as follows: Municipal land price data are obtained from official statistics published by the Ministry of Land, Infrastructure, Transport and Tourism. Tourism data are provided by the Japan Travel and Tourism Association, which estimates the annual number of domestic tourist arrivals in each municipality based on anonymized smartphone location information. Tourists are defined as domestic residents who travel at least 20 kilometers from their place of residence to visit a tourism destination, and the data are population-adjusted to improve representativeness. These statistics comprehensively cover various tourism destinations nationwide.

As mediating variables, the number of accommodation and food service establishments and their employees are included, based on municipal-level industrial statistics. To isolate the impact of tourism on land prices, a set of control variables is added to account for key demand- and supply-side fundamentals of local land markets. Specifically, we control for population size (proxying housing demand), labor force ratio (economic activity), total housing stock and vacant housing units (local housing supply conditions), and employment shares in the primary, secondary, and tertiary sectors (municipal economic structure). These controls are intended to remove confounding variation in land prices unrelated to tourism development. Most variables, except for labor force ratio, are log-transformed to reduce skewness and facilitate elasticity interpretation. Descriptive statistics for all variables are reported in Table 1. The dataset provides broad spatial and temporal coverage across Japan, supporting robust empirical analysis.

To clarify coverage differences across sources, the number of observations varies by variable in Table 1. Land prices are not reported for a small subset of municipalities or years, and housing stock and vacancy statistics have occasional gaps in the official releases. After merging all sources by municipality–year, observations that lack any variable required by a given specification are excluded from that specification only. We do not impute missing values. Accordingly, the econometric samples are constructed using listwise deletion within each model. The baseline and mediation analyses use the common set of municipalities–year observations with complete information on land prices, tourist arrivals, and controls, which yields an effective sample size of 4,750. Models with lagged regressors necessarily lose the first year per municipality, which reduces the effective sample size. The non-residential land price analysis relies on the subset of municipalities for which non-residential appraisals are reported, resulting in a smaller sample. These choices keep the identifying variation transparent while avoiding distributional assumptions about missing data.

Table 1. Descriptive Statistics of Main Variables

| Variable | N | Mean | Std. Dev. | Min | Max |
| --- | --- | --- | --- | --- | --- |
| Tourist Arrivals | 7519 | 772,478.34 | 1,943,732 | 26 | 35,707,778 |
| Land Price (yen) | 6080 | 107,383.07 | 372,119 | 1,750 | 8,614,345 |
| Population | 7484 | 66,402.49 | 98,001 | 156 | 920,372 |
| Labor Force Ratio | 7484 | 0.546 | 0.063 | 0.290 | 0.770 |
| Housing Units | 4818 | 51,601.18 | 59,918 | 1,974 | 544,978 |
| Vacant Housing Units | 4818 | 6,973.35 | 7,567 | 108 | 60,570 |
| Primary Sector Employment | 7520 | 1,035.18 | 1,197 | 0 | 10,917 |
| Secondary Sector Employment | 7520 | 6,976.04 | 9,522 | 0 | 92,389 |
| Tertiary Sector Employment | 7520 | 21,486.35 | 33,427 | 0 | 335,702 |
| Accommodation & Food Establish. | 7520 | 316.57 | 541 | 0 | 5,367 |
| Accommodation & Food Employment | 7520 | 2,470.14 | 5,007 | 0 | 68,167 |

Notes: All values are at the municipality-year level. All monetary amounts are in yen. Employment and housing statistics are reported as counts. Labor force ratio is unitless. N differs across variables due to coverage gaps and the source merge. No imputation is performed; samples are constructed via listwise deletion within each specification.

# 4. Methodology

This section outlines the empirical strategies used to investigate the impact of tourism development on land prices in Japanese municipalities. We describe the baseline regression approach, mediation analysis, threshold effect identification, heterogeneity analysis, and robustness checks.

## 4.1 Baseline Panel Fixed Effects Regression

To estimate the elasticity of land prices with respect to tourism activity, we begin with a baseline panel regression model that includes year fixed and municipality fixed effects to control for time-specific shocks affecting all municipalities simultaneously. The model is specified as follows:

$$log(LandPrice_{it}) = \alpha_0 + \beta_0 Log(Tourism_{it}) + \gamma_0 X_{it} + \mu_i + \lambda_t + \varepsilon_{it}$$

where $log(LandPrice_{it})$ is the log of land price in municipality $i$ at year $t$, $log(Tourism_{it})$ is the log of tourist arrivals, and $X_{it}$ is a vector of control variables (including population, labor force ratio, housing stock, vacant houses, employment structure, etc.). Municipality fixed effects ($\mu_i$) control for all time-invariant unobserved heterogeneity across municipalities. Year fixed effects ($\lambda_t$) absorb nationwide macroeconomic cycles, policy shocks, or other time-varying factors common to all municipalities. The disturbance term $\varepsilon_{it}$ captures idiosyncratic shocks not explained by the covariates and fixed effects; it is assumed to have zero conditional mean and finite variance given $X_{it}$, $\mu_i$, and $\lambda_t$, and standard errors are clustered at the municipality level to allow for heteroskedasticity and serial correlation within municipalities. What this specification cannot control for are municipality-specific time-varying unobservable and potential simultaneity between tourism and land prices. To mitigate these concerns, we later incorporate lag specifications and permutation-based placebo tests (Section 4.5), and the results remain robust. All continuous variables are log-transformed (except labor force ratio, which is retained in its original form) to interpret the coefficients as elasticities. This model structure leverages panel data advantages for causal identification, controlling for all time-invariant unobserved heterogeneity.

## 4.2 Mediation Analysis: Service Sector Pathways

To explore the mechanism through which tourism may affect land prices, we conduct a multi-step mediation analysis focusing on the service industry (specifically, the accommodation and restaurant sectors):

Step 1: Effect of Tourism on Service Sector Expansion

$$log(Mediator_{it}) = \alpha_1 + \beta_1 log(Tourism_{it}) + \gamma_1 X_{it} + \mu_i + \lambda_t + \varepsilon_{it}$$

Step 2: Effect of Both Tourism and Service Sector on Land Prices

$$log(LandPrice_{it}) = \alpha_2 + \beta_2 log(Tourism_{it}) + \delta log(Mediator_{it}) + \gamma_2 X_{it} + \mu_i + \lambda_t + \varepsilon_{it}$$

where $\ln(Mediator_{it})$ refers to either the log of the number of service sector establishments or the log of employees in those sectors. A significant coefficient $\delta$ suggests a mediating effect of the service industry in the tourism–land price relationship. The indirect effect is tested using the Sobel test or bootstrapping.

## 4.3 Threshold Effect Analysis: Hansen Model and Stepwise Grouping

To detect possible nonlinear or threshold effects in the tourism–land price relationship, we employ two complementary strategies:

**(a) Panel Threshold Regression (Hansen, 1999):** We implement the panel threshold regression method to endogenously estimate whether there is a critical level of tourism activity at which its impact on land prices shifts. $\theta^*$ is the estimated log-threshold value. The statistical significance of the threshold effect is evaluated using a likelihood ratio test with bootstrapped confidence intervals (Hansen, 1999, Journal of Econometrics). This approach allows us to endogenously detect a 'tipping point', the critical level of tourism above which land prices start to rise significantly, rather than assuming the effect is always the same.

$$log(LandPrice_{it}) = \begin{cases} \alpha_3 + \beta_3 \log(Tourism_{it}) + \gamma_3 X_{it} + \mu_i + \lambda_t + \varepsilon_{it}, if \ log(Tourism_{it}) \leq \theta^* \\ \alpha_4 + \beta_4 \log(Tourism_{it}) + \gamma_4 X_{it} + \mu_i + \lambda_t + \varepsilon_{it}, if \ log(Tourism_{it}) > \theta^* \end{cases}$$

**(b) Stepwise Grouping Analysis:** To complement the formal threshold estimation, we also conduct grouped regressions by partitioning the sample using the median, mean, tertiles, deciles, and fine-grained quantiles (e.g., 80th–97th percentiles). For each grouping, we estimate the model separately for municipalities below and above each cutoff and systematically compare the coefficients of log tourism to identify the exact regime where the effect becomes statistically significant. $g$ denotes the group. By comparing the estimated coefficients and their statistical significance across cutoffs, we empirically locate the exact quantile above which tourism's effect on land prices becomes statistically significant.

$$log(LandPrice_{it}) = \alpha_a + \beta_g \log(Tourism_{it}) + \gamma_g X_{it} + \mu_i + \lambda_t + \varepsilon_{it}$$

## 4.4 Heterogeneity Analysis: City Size and Regional Variation

To test H4, we examine whether the elasticity of land prices with respect to tourism differs across municipality types. We operationalize spatial heterogeneity along two dimensions, municipal population size and region, and re-estimate the baseline specification on the corresponding subsamples while keeping the same controls and fixed effects; only the estimation sample varies.

Municipalities are first split by population size into Small and Large, and the baseline model is estimated separately for each size group:

$$log(LandPrice_{it}) = \alpha_{size} + \beta_{size} \log(Tourism_{it}) + \gamma' X_{it} + \mu_i + \lambda_t + \varepsilon_{it}, \quad (i,t) \in S_{size}, size \in \{Small, Large\}$$

where $\beta_{size}$ is the group-specific elasticity for the given **size**. Group indicators are used only to form subsamples and are not entered as regressors because their main effects are absorbed by municipality fixed effects in the baseline specification. We report $\hat{\beta}_{Small}$ and $\hat{\beta}_{Large}$ and assess equality across groups with the Wald test of $H_0: \beta_{Small} = \beta_{Large}$.

We then assess regional heterogeneity by re-estimating the same specification within each region:

$$log(LandPrice_{it}) = \alpha_r + \beta_r \log(Tourism_{it}) + \gamma' X_{it} + \mu_i + \lambda_t + \varepsilon_{it}, \quad (i,t) \in S_r, r \in R$$

The parameter $\beta_r$ is the region-specific elasticity. We compare each $\beta_r$ with a reference region and conduct a joint Wald test of equality across regions. Regional indicators are time invariant and therefore not entered explicitly.

In all cases, standard errors are clustered at the municipality level, effective sample sizes differ across subsamples and estimates for very small cells are interpreted with caution. The results section reports $\{\hat{\beta}_{Small}, \hat{\beta}_{Large}\}$ with the p-value for equality, as well as the set $\{\hat{\beta}_r\}_{r \in R}$ and the joint regional test.

## 4.5 Robustness Checks

To assess the robustness of our findings within the same modeling framework, we conduct two complementary exercises. First, we probe timing using two lag structures. In the lagged-only specification we include $log(Tourism_{i,t-1})$ but exclude the contemporaneous term, whereas in the current+lagged specification we include both $log(Tourism_{it})$ and $log(Tourism_{i,t-1})$. This design tests whether the tourism effect on $log(LandPrice_{it})$ is contemporaneous, delayed, or both. Estimates are obtained on the same set of controls and fixed effects as in the baseline (the lagged models naturally drop the first year per municipality). The magnitude and significance of the tourism elasticity remain stable across these timing choices (see Table A1 in Appendix).

Second, we implement permutation-based placebo tests that respect the panel's dependence structure. Rather than permuting raw regressors or reordering time—which would violate exchangeability under serial correlation—we use a residual-permutation procedure. Specifically: (i) estimate a restricted model that excludes $log(Tourism_{it})$ but keeps the same fixed effects and controls; (ii) collect the fitted values and residuals $\hat{\varepsilon}_{it}$; (iii) within each municipality, randomly permute the residuals (as a robustness variant, we also apply cluster-wise sign flips), and form pseudo outcomes $\tilde{y}_{it} = \hat{y}_{it}^{\text{restricted}} + \hat{\varepsilon}_{it}^{\pi}$; (iv) re-estimate the full model on $\tilde{y}_{it}$ to obtain a placebo coefficient. Repeating steps (iii)–(iv) 1,000 times yields an empirical null distribution for the tourism coefficient under no effect. The observed coefficient lies in the extreme upper tail of this distribution (empirical p-values reported alongside the main tables), indicating that the baseline results are unlikely to be driven by spurious correlation or model artifacts.

Taken together, the timing exercise and residual-permutation placebo tests corroborate that the positive association between tourist arrivals and municipal land prices is not a statistical artifact and is robust to alternative timing assumptions and to inference those accounts for within-municipality dependence.

## 5. Results

### 5.1 Tourists and land price

Table 2 reports a set of nested fixed-effects specifications for the association between tourism and municipal land prices. Across all variants, the coefficient on log (Tourist Arrivals) is positive and statistically significant, ranging from 0.039 (fully controlled) to 0.048 (parsimonious). Interpreted as elasticities, a 10% increase in tourist arrivals is associated with a 0.39–0.48% increase in land prices within the same municipality over time, conditional on the common control set and fixed effects. The signs of the controls accord with priors: larger population and a greater housing stock are associated with higher land prices, whereas a higher share of primary/secondary employment correlates with lower prices. Model fit is high (within $R^2$ up to 0.723), and information criteria in the Appendix favor the preferred specification. Overall, these results provide clear evidence consistent with H1, indicating a stable and economically small-but-meaningful elasticity of land prices with respect to tourism in Japanese municipalities.

**Table 2. Stepwise Regression Results for Land Price**

|  | (1) | (2) | (3) | (4) | (5) | VIF |
|---|---|---|---|---|---|---|
| Dependent Var. | log (Land Price) | log (Land Price) | log (Land Price) | log (Land Price) | log (Land Price) |  |
| log (Tourist Arrivals) | 0.048*** | 0.042*** | 0.040*** | 0.039*** | 0.039*** | 1.02 |
|  | (0.005) | (0.005) | (0.005) | (0.006) | (0.006) |  |
| Population |  | 2.14e-6** | 2.19e-6** | 2.36e-6** | 2.31e-6** | 1.22 |
|  |  | (1.05e-6) | (1.08e-6) | (1.14e-6) | (1.16e-6) |  |
| Labor Force Ratio |  |  | 12.21*** | 11.92*** | 11.95*** | 1.21 |
|  |  |  | (0.21) | (0.21) | (0.21) |  |
| Total Housing Units |  |  |  | 9.31e-6*** | 8.86e-6*** | 1.51 |
|  |  |  |  | (1.45e-6) | (1.52e-6) |  |
| Vacant Housing |  |  |  | 1.08e-5*** | 1.00e-5*** | 1.30 |
|  |  |  |  | (3.58e-6) | (3.78e-6) |  |
| Primary Industry Employment |  |  |  |  | -0.0001*** | 1.15 |
|  |  |  |  |  | (7.11e-6) |  |
| Secondary Industry Employment |  |  |  |  | -3.03e-5*** | 1.21 |
|  |  |  |  |  | (2.47e-6) |  |
| Tertiary Industry Employment |  |  |  |  | -7.30e-6*** | 1.34 |
|  |  |  |  |  | (2.29e-6) |  |
| Year FE | Yes | Yes | Yes | Yes | Yes |  |
| Municipality FE | Yes | Yes | Yes | Yes | Yes |  |
| Constant | 3.99*** | 3.67*** | 3.61*** | 3.56*** | 3.56*** |  |
|  | (0.10) | (0.13) | (0.13) | (0.14) | (0.14) |  |
| N | 4750 | 4750 | 4750 | 4750 | 4750 |  |
| $R^2$ | 0.697 | 0.706 | 0.721 | 0.723 | 0.723 |  |

Notes: Standard errors in parentheses. ***p<0.01, **p<0.05, *p<0.1.

## 5.2 Mediators

We treat accommodation and food services (AFS) as the mediator. AFS is measured with two consistently observed scale indicators in the municipal panel: the number of AFS establishments and AFS employment. International statistical frameworks classify AFS as tourism-characteristic industries, and these variables are standard proxies for tourism-related capacity and jobs. Increases in these indicators should therefore be interpreted as expansion of local AFS capacity rather than changes in quality or productivity (OECD, 2017; OECD, 2024).

Within municipalities over time, tourist arrivals covary strongly with both mediators. As showed in Table 3, higher arrivals are associated with sizeable increases in the number of AFS establishments (0.112, $p < 0.001$) and AFS employment (0.117, $p < 0.001$). This pattern accords with international practice that views tourism as stimulating capacity deepening and labor absorption in accommodation and restaurants (Stacey, 2015; OECD, 2024).

Including both AFS mediators simultaneously in the land-price regression yields large and precisely estimated coefficients for establishments and employment, 0.483 and 0.457 respectively ($p < 0.001$) in table 4, while the coefficient on tourist arrivals becomes negative and significant. We read this attenuation as evidence consistent with a mechanism in which tourism raises land prices primarily insofar as it expands AFS capacity and employment, thereby increasing demand for commercial and related urban land uses. Following best practice in the mediation literature, we summarize the indirect effect with the product of coefficients and obtain uncertainty via the delta method or bootstrap (Imai, Keele, & Yamamoto, 2010). In keeping with cautious phrasing in top journals, we interpret these findings as consistent with the AFS-expansion channel rather than claiming complete mediation (Mas & Moretti, 2009).

Our mediators are scale-type proxies and do not capture all facets of sector development such as quality upgrading, average establishment size, or platform capacity. Their use nonetheless accords with established measurement practice, in which AFS accounts for a large share of direct tourism employment. The findings should therefore be read as consistent with an AFS-expansion channel, not as an exhaustive test of every pathway (OECD, 2017; OECD, 2024).

**Table 3. Effects of Tourist Arrivals on Mediators**

| Variables | log (Number of Establishments) | log (Number of Employees) |
| --- | --- | --- |
| | Coef. (Std. Err.) | Coef. (Std. Err.) |
| Log (Tourist Arrivals) | 0.112 *** (0.005) | 0.117 *** (0.006) |
| Population | 3.18e-6 *** (9.02e-7) | 3.64e-6 *** (1.02e-6) |
| Labor Force Ratio | 7.303 *** (0.154) | 8.108 *** (0.175) |
| Total Housing Units | 2.12e-6 * (1.12e-6) | 3.18e-6 ** (1.28e-6) |
| Vacant Housing | 2.92e-7 (2.80e-6) | 2.06e-6 (3.22e-6) |
| Primary Industry Employment | -7.69e-6 (5.28e-6) | -1.03e-5 * (6.08e-6) |
| Secondary Industry Employment | 1.99e-6 (1.83e-6) | 1.09e-6 (2.11e-6) |
| Tertiary Industry Employment | -2.88e-6 * (1.70e-6) | -2.90e-6 (1.95e-6) |
| Year FE | Yes | Yes |
| Municipality FE | Yes | Yes |
| Constant | 0.557 *** (0.109) | 0.540 *** (0.124) |
| N | 4750 | 4750 |
| $R^2$ | 0.765 | 0.774 |

**Table 4. Effects of Tourist Arrivals and Mediators on Land Price**

| Variables | log (Land Price) | log (Land Price) |
| --- | --- | --- |
| | (Establishments as Mediator) | (Employees as Mediator) |
| | Coef. (Std. Err.) | Coef. (Std. Err.) |
| Log (Tourist Arrivals) | -0.015 ** (0.005) | -0.015 ** (0.005) |
| Log (Number of Establishments) | 0.483 *** (0.014) | |
| Log (Number of Employees) | | 0.457 *** (0.012) |
| Population | 2.26e-6 ** (1.15e-6) | 2.33e-6 ** (1.15e-6) |
| Labor Force Ratio | 11.960 *** (0.211) | 11.971 *** (0.211) |
| Total Housing Units | 8.75e-6 *** (1.51e-6) | 8.83e-6 *** (1.52e-6) |
| Vacant Housing | 9.91e-6 *** (3.77e-6) | 9.99e-6 *** (3.77e-6) |
| Primary Industry Employment | -0.0001 *** (7.08e-6) | -0.0001 *** (7.08e-6) |
| Secondary Industry Employment | -3.02e-5 *** (2.46e-6) | -3.03e-5 *** (2.46e-6) |
| Tertiary Industry Employment | -7.24e-6 *** (2.28e-6) | -7.31e-6 *** (2.28e-6) |
| Year FE | Yes | Yes |
| Municipality FE | Yes | Yes |
| Constant | 3.563 *** (0.141) | 3.556 *** (0.141) |
| N | 4750 | 4750 |
| $R^2$ | 0.735 | 0.735 |

## 5.3 Threshold Effects: Nonlinearity in the Tourism–Land Price Relationship

To capture potential nonlinearities, we employ both stepwise grouping and panel threshold regression approaches, and the results are presented in Tables 5–7. Our analysis reveals a pronounced and statistically significant threshold effect in the tourism–land price relationship. Specifically, using Hansen's (1999) panel threshold regression, we estimate an optimal threshold of approximately 4,450,000 annual tourist arrivals per municipality (with a 95% confidence interval ranging from 4,290,000 to 4,620,000). Below this threshold, the effect of tourism on land prices is negligible and not statistically significant (coefficient = 0.011, t = 1.08, p > 0.1). However, once this threshold is surpassed, the coefficient sharply increases to 0.064 (t = 3.52, p < 0.01), indicating a strong and robust positive effect of tourism on land prices in high-tourism municipalities. This threshold effect is visually depicted in Figure 2 where the fitted relationship between tourist arrivals and municipal land prices demonstrates a marked increase in slope once the critical cutoff is exceeded.

Stepwise grouping regressions, partitioned by mean, median, tertiles, and higher quantiles, consistently support this result. They show that significant positive impacts only emerge in municipalities with tourism levels above the identified cutoff. For example, high group coefficients rise to 0.146–0.359 and are always highly significant. Figure 3 further illustrates this pattern by plotting the p-value trend for the low-tourism group across quantile thresholds, highlighting that statistical significance in the tourism–land price relationship emerges only in the uppermost quantiles of tourist arrivals. These findings confirm Hypothesis 3 and suggest that the positive effect of tourism on land prices is not universal but instead concentrated among destinations that attract large numbers of tourists. This nonlinear and regime-dependent effect has critical policy implications. Land price pressures, and thus risks of gentrification or affordability issues, are likely to be most acute in "superstar" tourist destinations, calling for targeted planning and intervention.

**Table 5. Threshold Effects of Tourist Arrivals on Land Price (Different Groupings)**

| Threshold | Group | Coefficient | Std. Error | p-value | Significance |
|---|---|---|---|---|---|
| Middle Threshold | Low | -0.001 | 0.0098 | 0.926 | |
| | High | 0.146 | 0.0151 | <0.001 | *** |
| Average Value Threshold | Low | 0.005 | 0.0102 | 0.624 | |
| | High | 0.146 | 0.0144 | <0.001 | *** |
| Trisection Threshold | Low | 0.011 | 0.0125 | 0.393 | |
| | Middle | -0.007 | 0.0444 | 0.873 | |
| | High | 0.199 | 0.0215 | <0.001 | *** |

**Table 6. Threshold Effects of Tourist Arrivals on Land Price at Different Top Quantiles**

| Threshold (Quantile, log value) | Group | Coefficient | Std. Error | p-value | Observations | Significance |
|---|---|---|---|---|---|---|
| 80% (14.06) | Low group | -0.002 | 0.0073 | 0.821 | 3800 | |
| | High group | 0.242 | 0.0326 | <0.001 | 950 | *** |
| 90% (14.68) | Low group | 0.004 | 0.0065 | 0.506 | 4275 | |
| | High group | 0.282 | 0.0548 | <0.001 | 475 | *** |
| 95% (15.14) | Low group | 0.014 | 0.0061 | 0.027 | 4512 | ** |
| | High group | 0.326 | 0.0960 | 0.001 | 238 | *** |
| 97% (15.52) | Low group | 0.019 | 0.0059 | 0.002 | 4607 | ** |
| | High group | 0.359 | 0.1400 | 0.011 | 143 | ** |

Notes: The "High group" is defined as log (Tourist Arrivals) above the given threshold; "Low group" is below or equal to the threshold.

**Table 7. Results of Panel Threshold Regression: Impact of Tourist Arrivals on Land Price**

| Group | Coefficient | t-value | Significance | 95% Confidence Interval | Threshold Value (Tourist Arrivals) |
|---|---|---|---|---|---|
| Below threshold (≤ 4,450,000) | 0.011 | 1.08 | Not significant | [-0.006, 0.028] | 4,450,000 |
| Above threshold (> 4,450,000) | 0.064 | 3.52 | *** | [0.029, 0.099] | 4,450,000 |

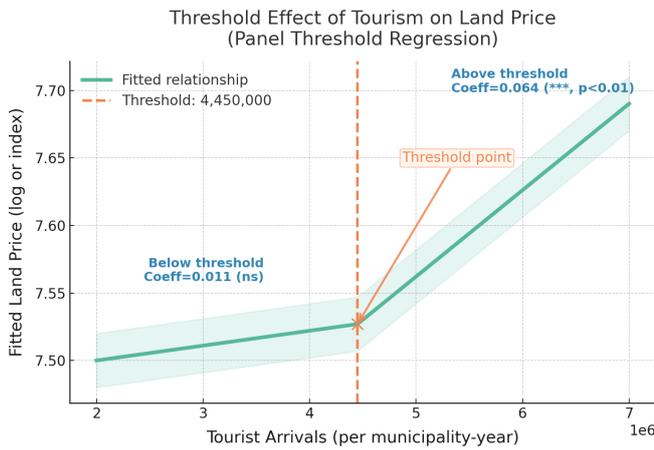

Figure 2. Panel Threshold Regression

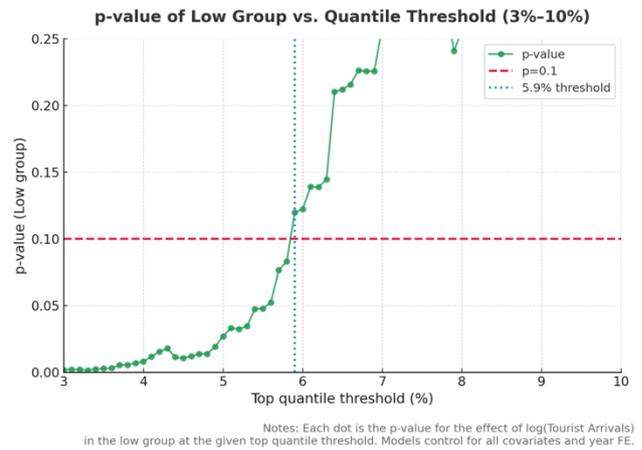

Figure 3. P-value trend of Low Groups

## 5.4 Heterogeneity Analysis: Variation by Municipality Size

To test H4, we assess how the tourism–land price elasticity varies across municipality types and intensity regimes. Using subsample regressions by population size, we find a sharp pattern (table 8). Among large municipalities, the elasticity of $log(LandPrice)$ with respect to $log(Tourist\ Arrivals)$ is positive and statistically significant in the main regression (coefficient =0.0211, p=0.006), and becomes markedly larger in the high-tourism subgroup (coefficient =0.331, p<0.01). By contrast, large cities with low tourist inflows exhibit a negative and significant association, plausibly consistent with displacement or saturation effects. Among small municipalities, the overall association is not statistically significant and can turn negative in the highest tourism quantiles (e.g., coefficient =−0.598, p<0.05). While estimates in extreme cells are inevitably less precise, the cross-group pattern is consistent: the elasticity is concentrated in large, tourism-intensive places and is weak or even adverse elsewhere. All subsample specifications retain the same controls and fixed effects as the baseline, and standard errors are clustered at the municipality level.

**Table 8. Heterogeneity by City Size: Main and Threshold Regression Results**

| Group | Regression Type | Coefficient | Std. Error | p-value | N | Significance |
|---|---|---|---|---|---|---|
| Large | Main | 0.0211 | 0.0077 | 0.006 | 2376 | *** |
|  | Low group (<5.9%) | -0.018 | 0.0084 | 0.033 | 2213 | ** |
|  | High group (>5.9%) | 0.331 | 0.1083 | 0.003 | 163 | *** |
| Small | Main | -0.0054 | 0.0071 | 0.451 | 2374 |  |
|  | Low group (<5.9%) | -0.0064 | 0.0074 | 0.383 | 2342 |  |
|  | High group (>5.9%) | -0.598 | 0.2674 | 0.037 | 32 | ** |

## 5.5 Non-Residential Land Price Analysis and Tourism Impact

The results show that while tourism has a positive and significant effect on non-residential land prices (Table 9), the magnitude of this impact is lower than that for overall land prices. The estimated elasticity for non-residential land is 0.028, compared to 0.039 for total land prices, indicating that non-residential land price growth associated with tourism is less pronounced.

A possible explanation is that tourism-driven growth in service sector employment primarily increases demand for housing, which pushes up residential land prices more than non-residential prices. This difference suggests the need for further research to clarify how tourism affects various land market segments and to examine the underlying mechanisms in greater detail.

**Table 9. Non-Residential Land Price Regression Results**

| Variable | Coef. | Std.Err | t | p-value | 95% Conf. Interval |
|---|---|---|---|---|---|
| Log (Tourist Arrivals) | 0.0278** | 0.0084 | 3.29 | 0.001 | [0.011, 0.044] |
| Population | 5.13e-7 | 1.64e-6 | 0.31 | 0.754 | [-2.71e-6, 3.73e-6] |
| Labor Force Participation | 10.83*** | 0.329 | 32.95 | 0.000 | [10.19, 11.46] |
| Total Housing Units | 5.54e-6** | 2.17e-6 | 2.56 | 0.011 | [1.28e-6, 9.80e-6] |
| Vacant Units | 7.80e-6 | 5.41e-6 | 1.44 | 0.150 | [-2.80e-6, 1.84e-5] |
| Primary Sector Employment | -0.0001*** | 1.02e-5 | -10.44 | 0.000 | [-0.00013, -8.07e-5] |
| Secondary Sector Employment | -2.81e-5*** | 3.50e-6 | -8.04 | 0.000 | [-3.50e-5, -2.12e-5] |
| Tertiary Sector Employment | -2.35e-6 | 3.39e-6 | -0.69 | 0.488 | [-8.99e-6, 4.30e-6] |
| Year FE | YES | | | | |
| Municipality FE | YES | | | | |
| Observations | 3765 | | | | |
| $R^2$ | 0.64 | | | | |

# 6. Discussion and Limitations

## 6.1 Discussion

This study provides clear nationwide evidence that the relationship between tourism and municipal land prices in Japan is characterized by marked nonlinearity and strong spatial variation. Our empirical results, visually summarized in Figure 4, demonstrate that significant land price appreciation is highly concentrated in a very small group of "superstar" cities—specifically, those ranking in the top 5.9% for annual tourist arrivals. In these cities, the estimated elasticity reaches 0.33 and is robust across specifications. This sharply contrasts with the experience of most other municipalities, where even moderate or above-average growth in tourism is not associated with meaningful changes in land prices. Importantly, our analysis reveals that in smaller cities with high tourism intensity, the effect of tourism on land prices can even become negative, suggesting that the benefits of tourism are not only uneven but may also bring unexpected risks, such as displacement of local demand or resource strain.

The observed heterogeneity fundamentally challenges the prevailing "one-size-fits-all" approach to tourism-led urban development. It becomes clear that tourism does not universally drive-up land values; instead, the local context—including city size, economic structure, and the scale of tourism—plays a decisive role. This insight not only refines theoretical expectations but also provides new evidence for international debates on the risks and opportunities associated with tourism-driven urban change (Chiu & Yeh, 2017; Li et al., 2022).

Delving into the mechanisms, our mediation analysis shows that the expansion of the accommodation and food service sectors almost fully accounts for the observed land price effects. Rather than simple increases in tourist numbers, it is the sustained transformation and upgrading of local service economies that amplifies both the positive and negative impacts of tourism on the urban land market. Where sectoral growth is strong, it can drive job creation and local income gains, but it may also intensify gentrification pressures and worsen housing affordability, particularly in high-demand tourist cities. This aligns with recent findings that highlight the dual-edged nature of tourism's impact: new opportunities for local development on the one hand, and heightened risks of social and economic exclusion on the other (Biagi et al., 2015; Gotham, 2005; Yoshida & Kato, 2024).

Our results further reveal that the impact of tourism is not limited to residential markets, although the effects are clearly strongest there. The relatively small and statistically marginal impact on non-residential land prices suggests that tourism-driven demand is more likely to exacerbate competition for housing, especially among workers and service sector employees. This can create additional affordability challenges, while the effect on commercial land is often diluted by a broader range of economic and policy factors (Paramati & Roca, 2019). Such findings underscore the need for a more nuanced understanding of how tourism interacts with different segments of urban land markets, and for closer integration between urban, housing, and tourism policy agendas.

The findings have direct implications for policy. For most municipalities, tourism-related land price pressures are negligible, which means that local governments can prioritize leveraging positive spillovers from tourism through investment in service sector innovation and infrastructure, without concern for widespread housing market disruption. However, in "superstar" destinations where both demand and price pressures are high, targeted policies are urgently needed. These may include expanding affordable housing programs, regulating the growth of short-term rentals, and adopting progressive property tax measures to mitigate affordability risks. Given the potential for price escalation to spill over into neighboring regions, metropolitan-level coordination on land use, housing supply, and rental regulation is likely to become increasingly important in areas such as the Kyoto–Osaka corridor.

At a broader level, the study's methodology strengthens confidence in these conclusions. By integrating panel threshold regression, mediation analysis, and extensive robustness checks, we provide a rigorous empirical foundation that addresses longstanding questions about the scale, mechanism, and spatial heterogeneity of tourism's impact on urban land markets.

It is also essential to recognize the limitations inherent in the current data and approach, which are discussed further in Section 6.2. For instance, while our analysis uncovers robust patterns at the municipal level, the limited time window and lack of micro-level data constrain our ability to fully identify displacement mechanisms or track welfare impacts for specific groups. Future research would benefit from more granular longitudinal data, including transaction-level housing records and household income surveys, to better understand who actually gains or loses from tourism-driven land price changes. Furthermore, our sample size for certain subgroups, such as extremely small or very high-tourism cities, is limited, and caution is warranted in interpreting these findings as definitive.

In summary, this study challenges prevailing assumptions about tourism and land prices by demonstrating that the effects are highly selective, deeply context-dependent, and shaped by both economic structure and the transformation of local service sectors. Effective policy must move beyond uniform national strategies, instead embracing a place-based, adaptive approach that accounts for the diversity of urban experiences in Japan and, by extension, in other rapidly changing tourism destinations worldwide.

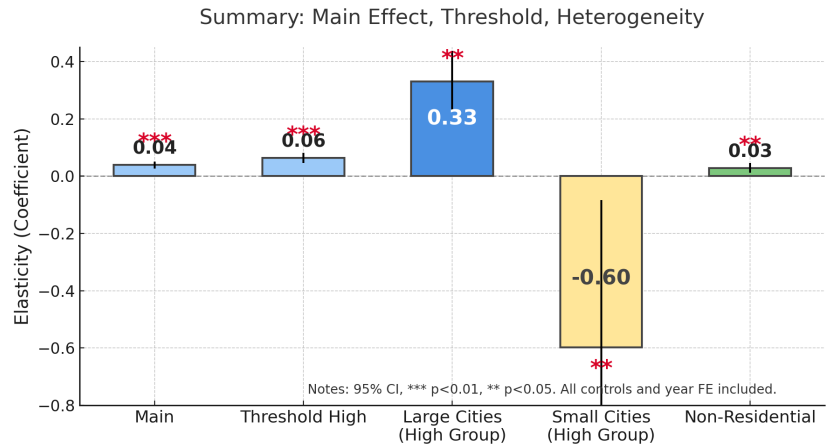

**Figure 4. Summary for this research**

## 6.2 Limitations

This study, while offering new empirical insights, has several important limitations that should be acknowledged. First, the relatively short time frame of the panel (2021–2024) constrains the ability to capture long-term or lagged effects, especially for structural transformation and gentrification, which typically unfold over a longer period (Mikulić et al., 2021; Vuković et al., 2023).

Second, the analysis relies exclusively on annual tourist arrivals as the measure of tourism activity, without directly considering tourist expenditure or consumption patterns. It is actual spending rather than visitor numbers alone that fundamentally drives local economic growth, supports business revenues, and determines the broader welfare effects of tourism. By focusing only on arrivals, this study may overlook substantial variation in per capita spending, trip purpose, or length of stay, as well as important differences between volume-driven and value-driven impacts. Future research should seek to incorporate more granular data on tourist spending to clarify how different types and levels of tourism affect land markets and community outcomes.

Third, while the mediation role of the accommodation and food service sectors is empirically established, other possible mechanisms such as infrastructure investment, the proliferation of short-term rentals, and local regulatory shifts could not be examined due to data constraints (Vizek et al., 2024). The absence of transaction-level, migration, or rental data also prevents direct identification of patterns of displacement or a full understanding of the pathways through which tourism affects land values (Gotham, 2005; Yoshida & Kato, 2024).

Fourth, issues of endogeneity remain. Despite employing fixed effects, lagged variables, and placebo tests, it is difficult to rule out reverse causality or omitted variable bias. For example, municipalities with rising land prices may subsequently invest more in tourism or attract further development (Pulido-Fernández & Cárdenas-García, 2021). The lack of strong instrumental variables reflects a broader empirical challenge in this research area.

Fifth, the use of municipality-level data, though suitable for national-scale analysis, may mask important intra-city disparities and local heterogeneity (Cong et al., 2025). Without more granular micro-level data, the effects of tourism-induced land price changes on different neighborhoods or population groups remain unclear.

Finally, a central and policy-relevant question remains unresolved: Who ultimately benefits from tourism-driven land price appreciation, and through what mechanisms are these gains distributed? While our findings suggest that residential land prices tend to rise more than non-residential ones, indicating a key role for increased housing demand, this distinction is far from definitive. Without detailed transaction records or information on buyers' identities and the intended use of properties, it is impossible to determine whether the gains primarily accrue to local residents, outside investors, or other stakeholders. Even with such data, distinguishing genuine improvements in residents' well-being from increases driven by speculative or external forces would require more sophisticated analytical strategies. This question is examined further in Section 6.3, which explores the mechanisms and fiscal implications of tourism-led land price growth.

## 6.3 Mechanism and Fiscal Analysis

To better understand who benefits from the increase in tourist arrivals, we analyzed the impact of tourism growth on local tax revenue. As shown in Table 10, the results indicate a strong and positive link between tourism and fiscal outcomes. When tourist arrivals increase by 1%, local tax revenue rises by about 0.07%. This relationship remains robust even after accounting for differences in population size, employment structure, and housing stock. These findings suggest that tourism expansion is closely connected to economic development and improved municipal finances.

However, this result does not mean that every resident shares these benefits equally. The main reason for this uncertainty lies in the aggregate nature of the tax data. Local tax revenue combines payments from residents, businesses, and property owners, making it impossible to determine whether the economic gains are broadly shared within the community or concentrated among groups. In other words, while tourism clearly boosts overall economic activity and government revenue, we do not know how these benefits are distributed among different stakeholders.

This uncertainty highlights the complexity of how tourism affects land prices and local welfare. Figure 5 presents a conceptual framework that summarizes several possible pathways by which tourism may lead to higher land prices. One possible pathway is that increased tourism stimulates local economic growth, raises residents' incomes, and enables more people to afford better-quality housing. In this situation, land price appreciation reflects improvements in local prosperity and living standards, and the benefits are more likely to be widely shared.

However, Figure 5 also includes several other pathways, most of which primarily benefit nonlocal groups. For example, tourism may attract new workers, bring in outside investors, or encourage speculative activity. In these cases, the increase in demand for housing is driven by newcomers or external capital rather than rising local incomes. Land prices may rise, but the gains often go to property owners or investors, while ordinary households may not see improved welfare. Instead, they may face greater competition for limited housing, higher living costs, and increased affordability pressures.

It is important to recognize that this framework is based on previous research and theoretical reasoning, rather than being fully confirmed by our current data. Our empirical results support the idea that tourism stimulates local economies and boosts fiscal revenue, but they do not allow us to determine which specific mechanism is most important in practice. Whether land price increases signal genuine prosperity or increased burdens for residents depends on which pathway dominates.

Distinguishing between "prosperity-driven" and "burden-driven" land price increases is essential for designing effective policies. If higher prices reflect broad-based economic gains, tourism can contribute to sustainable and inclusive growth. If, however, price growth is mainly driven by outside demand or speculative investment, the resulting burdens may outweigh the benefits for local households. Future research should use more detailed data—such as household income surveys, property transaction records, or direct measures of housing quality—to clarify who benefits from tourism-led land price growth and to guide policy responses that ensure the positive effects of tourism are both widespread and sustainable.

**Table 10. Effect of Tourist Arrivals on Local Tax Revenue**

| Variable | Coefficient | Std. Error | t-value | p-value |
|---|---|---|---|---|
| **Log (Tourist Arrivals)** | 0.0721*** | 0.0050 | 14.33 | <0.001 |
| Population | 2.06e-5*** | 1.59e-6 | 13.00 | <0.001 |
| Labor Force Ratio | 8.58*** | 0.19 | 44.63 | <0.001 |
| Number of Dwellings | -2.79e-5*** | 2.71e-6 | -10.30 | <0.001 |
| Number of Vacant Houses | 6.29e-5*** | 4.89e-6 | 12.89 | <0.001 |
| Employment in Primary Sector | 3.90e-5*** | 6.04e-6 | 6.46 | <0.001 |
| Employment in Secondary Sector | -3.50e-6 | 2.39e-6 | -1.46 | 0.145 |
| Employment in Tertiary Sector | -1.22e-5*** | 2.49e-6 | -4.90 | <0.001 |
| Accommodation & Food Service Establishments | 0.00020* | 0.00008 | 2.07 | 0.038 |
| Employees in Accommodation & Food Service | -9.40e-6 | 1.19e-5 | -0.80 | 0.423 |
| Year FE | YES | | | |
| Municipality FE | YES | | | |
| Constant | 9.30*** | 0.13 | 71.71 | <0.001 |
| N | 3,014 | | | |
| R² | 0.851 | | | |

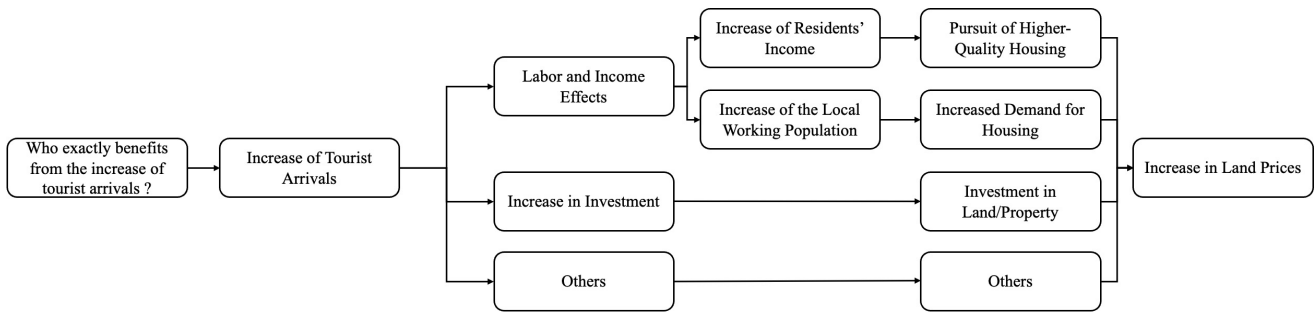

**Figure 5. Mechanisms Linking Tourist Arrivals to Land Price Increases: Who Benefits?**

# 7. Policy Implications

Based on the threshold and heterogeneity results identified in this study (Figure 4), several actionable policy lessons emerge for urban planners and policymakers. Most fundamentally, the existence of clear nonlinearities and spatial heterogeneity in tourism's impact on land markets calls for differentiated, place-based strategies rather than uniform national policies. For many municipalities that remain below the identified tourism threshold, moderate visitor growth does not pose a significant threat to land prices or housing affordability. In these contexts, policy should focus on harnessing positive economic spillovers by supporting local entrepreneurship, promoting service sector innovation, and investing in foundational infrastructure, as depicted on the left side of Figure 4.

In contrast, the right side of Figure 4 highlights the specific policy interventions needed for "superstar" tourist cities that exceed the threshold and face pronounced risks of land price escalation and gentrification. For these high-impact destinations, our results suggest the need for progressive property taxation to curb speculative investment, robust regulation of short-term rentals to protect residential supply, and the targeted expansion of affordable housing initiatives. Furthermore, given that the accommodation and food service industries are the main conduit for tourism's impact on land prices, integrated support and regulation of these sectors can help amplify positive effects while mitigating negative externalities.

Looking forward, the roadmap illustrated in Figure 4 underscores the necessity of adaptive, data-driven management. Since tourism flows and land market dynamics can shift rapidly in response to global shocks or domestic developments, ongoing monitoring and flexible policy adjustment are vital. Regional and national authorities can play a supportive role by investing in high-quality tourism and housing market data systems, providing technical guidance, and facilitating inter-municipal coordination to address spillover effects and emerging challenges. By explicitly referencing the steps outlined in Figure 6, local governments can better anticipate the impacts of tourism growth and craft tailored, responsive strategies that promote sustainable and inclusive urban development.

In summary, these findings reaffirm that no one-size-fits-all policy exists for managing the land market impacts of tourism. Instead, effective governance depends on nuanced, place-based strategies that are responsive to local economic structures, population dynamics, and the evolving spatial footprint of tourism. These insights are not only relevant for Japan but also provide a valuable reference for international policymakers seeking to balance economic growth, affordability, and urban resilience in the face of rising tourism.

Ultimately, no single policy can address the diverse impacts of tourism on land markets. Rather, adaptive and context-sensitive strategies, grounded in robust empirical evidence, are essential to balance growth, affordability, and social inclusion.

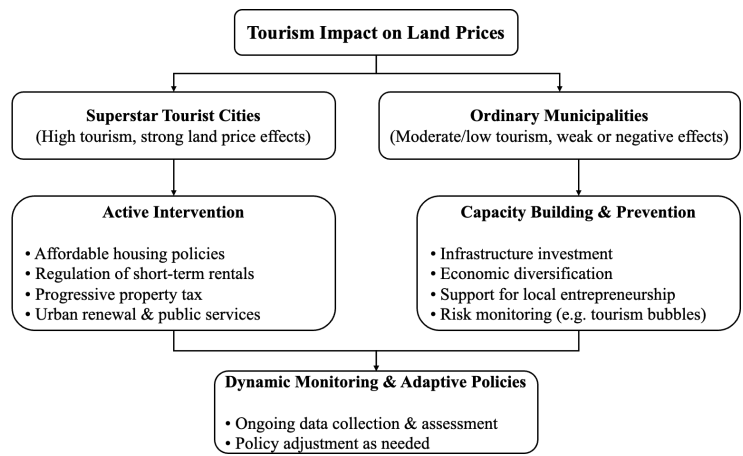

**Figure 6. Policy Roadmap**

# 8. Conclusion

This study demonstrates that the effects of tourism on urban land prices in Japan are far from universal or linear. Our nationwide analysis shows that land price appreciation driven by tourism is highly selective, with significant effects concentrated almost exclusively in a small group of "superstar" cities—those with the highest levels of tourist arrivals. For many municipalities, even those with moderate or above-average tourism, the impact on land prices is negligible or statistically insignificant. Notably, some small cities with intense tourism activity may even experience negative land price effects, highlighting that tourism does not guarantee widespread economic benefits and can also introduce new risks and pressures.

A central contribution of this study is the identification of the expansion of accommodation and food service sectors as the principal mechanism linking tourism to land price changes. It is not simply tourist numbers that matter, but the way local economies transform in response to tourism demand. Where sectoral upgrading is robust, positive spillovers on employment and income can occur. Yet these changes can also intensify affordability challenges and gentrification risks, especially in destinations that attract both large numbers of visitors and incoming workers.

The empirical evidence further points to pronounced spatial variation in the tourism–land price relationship. The magnitude and direction of tourism's impact are strongly dependent on local economic structure, city size, and the scale of tourism activity. Residential land markets are particularly sensitive, while the effects on non-residential land are weaker and less consistent. These patterns make clear that uniform, national-level policies are unlikely to be effective. Instead, adaptive and place-based policy responses are needed. In high-tourism cities, measures such as affordable housing expansion, regulation of short-term rentals, and targeted property taxation can help address mounting affordability pressures. In other municipalities, efforts to foster economic resilience and innovation in the service sector may be more appropriate.

It is also important to recognize that rising land prices are not inherently detrimental. When matched by improvements in local income and public services, price growth may indicate genuine prosperity. However, if land price escalation primarily benefits property owners or investors, it can undermine affordability and widen inequality. Understanding the underlying drivers and distributional consequences of land price changes remains a crucial challenge for both scholars and policymakers.

This study advances the literature by integrating panel threshold modeling, mediation analysis, and rigorous robustness checks, providing new clarity on the complex, nonlinear, and context-dependent relationship between tourism and land values. Nonetheless, limitations remain, including the lack of transaction-level and household data, and the relatively short observation window. Future research should focus on more granular, longitudinal datasets to clarify who ultimately benefits from or bears the costs of tourism-driven urban change.

Overall, our findings argue for moving beyond one-size-fits-all assumptions in the analysis and governance of tourism and land markets. As global tourism continues to reshape cities, lessons from the Japanese experience underscore the need to balance economic growth, housing affordability, and inclusive urban development, both nationally and internationally.

# Appendix A. Robustness and Spatial Visualization

**A.1 Robustness checks**

To assess the credibility and stability of the baseline estimates, we conduct two exercises using Table A1 and Figure A1. First, we augment the specification with lagged values of tourist arrivals to allow for delayed responses and to probe potential endogeneity. Both the contemporaneous and lagged coefficients remain positive and statistically significant, and the magnitudes are stable when the terms enter jointly. This pattern indicates that the relationship is not driven by transitory shocks and reduces concern about short run reverse causality.

Second, we implement a permutation placebo test as described in Section 4.5. We randomly reshuffle the treatment assignment and the time sequence one thousand times, record the estimated coefficients, and form a placebo distribution under the null of no effect. The baseline estimate lies in the extreme upper tail of this distribution, beyond the ninety five percent critical band, while the placebo distribution is tightly centered near zero. This result is consistent with the view that the findings are not artifacts of model choice, omitted variables, or spurious correlation. Taken together, these checks support a causal interpretation of the tourism–land price elasticity in Japan.

**A.2 Spatial visualization of the tourism component**

To make the heterogeneity transparent in geographic space, Figure A2 visualizes the tourism related component of predicted land prices implied by the panel estimates. For each municipality we compute the partial predictions attributable to tourism using the group specific elasticities reported in the main text. We then interpolate the municipality values to a continuous surface with a Gaussian spatial field using the SPDE framework in sdmTMB (Anderson et al., 2024). This procedure is used only for interpolation and presentation and does not re-estimate coefficients or alter identification.

The resulting map shows a marked concentration of high tourism related predicted values in the Tokyo, Osaka, and Nagoya metropolitan areas and a small number of well-known destinations, with much lower predicted values across most rural and peripheral municipalities. The spatial clustering mirrors the subsample, and threshold results and supports H4. The evidence cautions against uniform policies because benefits appear concentrated in a limited set of highly competitive cities, while most municipalities exhibit little measurable appreciation linked to tourism.

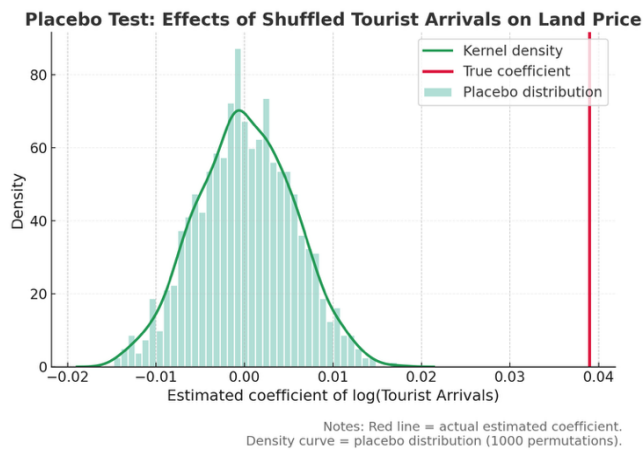

**Figure A1. Placebo Test**

**Table A1. Lagged Effects of Tourist Arrivals on Land Price**

| Variables | (1) Lagged Tourist Arrivals Only | (2) Both Current & Lagged Tourist Arrivals |
|---|---|---|
| | Coefficient (Std. Error) | Coefficient (Std. Error) |
| Log (Tourist Arrivals) | | 0.022 ** (0.007) |
| Log (Tourist Arrivals) t−1 | 0.041 *** (0.007) | 0.032 *** (0.007) |
| Population | 2.70e-6 ** (1.35e-6) | 2.93e-6 ** (1.35e-6) |
| Labor Force Ratio | 12.012 *** (0.239) | 12.010 *** (0.239) |
| Total Housing Units | 8.24e-6 *** (1.74e-6) | 7.75e-6 *** (1.75e-6) |
| Vacant Housing | 9.53e-6 ** (4.27e-6) | 1.01e-5 ** (4.27e-6) |
| Primary Industry Employment | -0.0001 *** (8.2e-6) | -0.0001 *** (8.2e-6) |
| Secondary Industry Employment | -3.14e-5 *** (2.84e-6) | -3.20e-5 *** (2.84e-6) |
| Tertiary Industry Employment | -7.30e-6 *** (2.64e-6) | -7.22e-6 *** (2.64e-6) |
| Year FE | Yes | Yes |
| Municipality FE | Yes | Yes |
| Constant | 3.520 *** (0.425) | 3.382 *** (0.426) |
| N | 3563 | 3563 |
| $R^2$ | 0.729 | 0.730 |

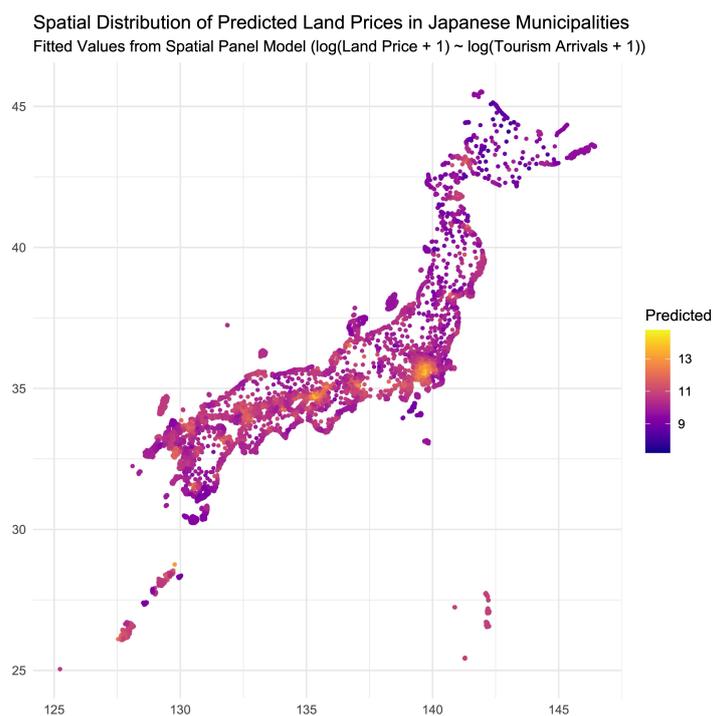

**Figure A2. Predicted land prices in Japanese municipalities, 2021–2024**

OECD et al. (2017), Tourism Satellite Account: Recommended Methodological Framework 2008, OECD Publishing, Paris, https://doi.org/10.1787/9789264274105-en.

Paramati, S. R., & Roca, E. (2019). Does tourism drive house prices in the OECD economies? Evidence from augmented mean group estimator. Tourism Management, 74, 392-395.

Pulido-Fernández, J. I., & Cárdenas-García, P. J. (2021). Analyzing the bidirectional relationship between tourism growth and economic development. Journal of Travel Research, 60(3), 583-602.

Song, C., Yin, T., Zhi, Q., Gu, J., & Li, X. (2025). How does tourism development affect the land market? Evidence from land transaction data in China. Tourism Review, 80(6), 1277-1299.

Stacey, J. (2015). Supporting quality jobs in tourism.

Vizek, M., Barbić, T., & Časni, A. Č. (2024). The impact of the tourism accommodation composition on housing prices: The case of Croatia. Tourism Economics, 30(1), 267-274.

Vuković, D. B., Maiti, M., & Petrović, M. D. (2023). Tourism employment and economic growth: Dynamic panel threshold analysis. Mathematics, 11(5), 1112.

Yoshida, M., & Kato, H. (2023). Housing affordability risk and tourism gentrification in Kyoto City. Sustainability, 16(1), 309.

Zhang, J. (2024). Tourism development and housing price: an interplay. Tourism Economics, 30(5), 1095-1114.
24